\documentclass[aps,twocolumn,showpacs,superscriptaddress,floatfix]{revtex4}

\usepackage[dvips]{graphicx}
\usepackage{xspace}
\usepackage{latexsym}
\usepackage{amssymb}
\begin{document}
%\begin{frontmatter}
\title{Search for Neutral Q-balls in Super-Kamiokande II}

\newcommand{\icrr}{\affiliation{Kamioka Observatory, Institute for Cosmic Ray Research, University of Tokyo, Kamioka, Gifu, 506-1205, Japan}}
\newcommand{\ncen}{\affiliation{Research Center for Cosmic Neutrinos, Institute for Cosmic Ray Research, University of Tokyo, Kashiwa, Chiba 277-8582, Japan}}
\newcommand{\bu}{\affiliation{Department of Physics, Boston University, Boston, MA 02215, USA}}
\newcommand{\bnl}{\affiliation{Physics Department, Brookhaven National Laboratory, Upton, NY 11973, USA}}
\newcommand{\uci}{\affiliation{Department of Physics and Astronomy, University of California, Irvine, Irvine, CA 92697-4575, USA}}
\newcommand{\csu}{\affiliation{Department of Physics, California State University, Dominguez Hills, Carson, CA 90747, USA}}
\newcommand{\cnu}{\affiliation{Department of Physics, Chonnam National University, Kwangju 500-757, Korea}}
\newcommand{\duke}{\affiliation{Department of Physics, Duke University, Durham, NC 27708 USA}}
\newcommand{\gmu}{\affiliation{Department of Physics, George Mason University, Fairfax, VA 22030, USA}}
\newcommand{\gifu}{\affiliation{Department of Physics, Gifu University, Gifu, Gifu 501-1193, Japan}}
\newcommand{\uh}{\affiliation{Department of Physics and Astronomy, University of Hawaii, Honolulu, HI 96822, USA}}
\newcommand{\ui}{\affiliation{Department of Physics, Indiana University, Bloomington,  IN 47405-7105, USA} }
\newcommand{\kek}{\affiliation{High Energy Accelerator Research Organization (KEK), Tsukuba, Ibaraki 305-0801, Japan}}
\newcommand{\kobe}{\affiliation{Department of Physics, Kobe University, Kobe, Hyogo 657-8501, Japan}}
\newcommand{\kyoto}{\affiliation{Department of Physics, Kyoto University, Kyoto 606-8502, Japan}}
\newcommand{\lanl}{\affiliation{Physics Division, P-23, Los Alamos National Laboratory, Los Alamos, NM 87544, USA}}
\newcommand{\lsu}{\affiliation{Department of Physics and Astronomy, Louisiana State University, Baton Rouge, LA 70803, USA}}
\newcommand{\umd}{\affiliation{Department of Physics, University of Maryland, College Park, MD 20742, USA}}
\newcommand{\MIT}{\affiliation{Department of Physics, Massachusetts Institute of Technology, Cambridge, MA 02139, USA}}
\newcommand{\duluth}{\affiliation{Department of Physics, University of Minnesota, Duluth, MN 55812-2496, USA}}
\newcommand{\miyagi}{\affiliation{Department of Physics, Miyagi University of Education, Sendai,Miyagi 980-0845, Japan}}
\newcommand{\suny}{\affiliation{Department of Physics and Astronomy, State University of New York, Stony Brook, NY 11794-3800, USA}}
\newcommand{\nagoya}{\affiliation{Department of Physics, Nagoya University, Nagoya, Aichi 464-8602, Japan}}
\newcommand{\nagoyaste}{\affiliation{Solar-Terrestrial Environment Laboratory, Nagoya University, Nagoya, Aichi 464-8601, Japan}}
\newcommand{\niigata}{\affiliation{Department of Physics, Niigata University, Niigata, Niigata 950-2181, Japan}}
\newcommand{\osaka}{\affiliation{Department of Physics, Osaka University, Toyonaka, Osaka 560-0043, Japan}}
\newcommand{\okayama}{\affiliation{Department of Physics, Okayama University, Okayama, Okayama 700-8530, Japan}}
\newcommand{\seoul}{\affiliation{Department of Physics, Seoul National University, Seoul 151-742, Korea}}
\newcommand{\shizuokaseika}{\affiliation{International and Cultural Studies, Shizuoka Seika College, Yaizu, Shizuoka 425-8611, Japan}}
\newcommand{\shizuokafukushi}{\affiliation{Department of Informatics in
Social Welfare, Shizuoka University of Welfare, Yaizu, Shizuoka 425-8611, Japan}}
\newcommand{\shizuoka}{\affiliation{Department of Systems Engineering, Shizuoka University, Hamamatsu, Shizuoka 432-8561, Japan}}
\newcommand{\skku}{\affiliation{Department of Physics, Sungkyunkwan University, Suwon 440-746, Korea}}
\newcommand{\tohoku}{\affiliation{Research Center for Neutrino Science, Tohoku University, Sendai, Miyagi 980-8578, Japan}}
\newcommand{\tokyo}{\affiliation{University of Tokyo, Tokyo 113-0033, Japan}}
\newcommand{\tokai}{\affiliation{Department of Physics, Tokai University, Hiratsuka, Kanagawa 259-1292, Japan}}
\newcommand{\tit}{\affiliation{Department of Physics, Tokyo Institute for Technology, Meguro, Tokyo 152-8551, Japan}}
\newcommand{\tsinghua}{\affiliation{Department of Engineering Physics, Tsinghua University, Beijing 100084, China}}
\newcommand{\warsaw}{\affiliation{Institute of Experimental Physics, Warsaw University, 00-681 Warsaw, Poland}}
\newcommand{\uw}{\affiliation{Department of Physics, University of Washington, Seattle, WA 98195-1560, USA}}
\newcommand{\tsukubanow}{\altaffiliation{ Present address: Department of Physics, Univ. of Tsukuba, Tsukuba, Ibaraki 305 8577, Japan}}
\newcommand{\okayamanow}{\altaffiliation{ Present address: Department of Physics, Okayama University, Okayama 700-8530, Japan}}
\newcommand{\marylandnow}{\altaffiliation{ Present address: University of Maryland School of Medicine, Baltimore, MD 21201, USA}}
\newcommand{\triunfnow}{\altaffiliation{ Present address: TRIUMF, Vancouver, British Columbia V6T 2A3, Canada}}
\newcommand{\icrrnow}{\altaffiliation{ Present address: Kamioka Observatory, Institute for Cosmic Ray Research, University of Tokyo, Kamioka, Gifu, 506-1205, Japan}}
\newcommand{\pennnow}{\altaffiliation{ Present address: Center for Gravitational Wave Physics, Pennsylvania State University, University Park, PA 16802, USA}}
%%%%%%%%%%%%%%%%%%%%%%%%%%%%%%%%%%%%%%%%%%%%%%%%%%%%%%%%%%%%%%%%%%%%%%%%%
%------------------------------------------------------------------------
%First authors
\author{Y.Takenaga}\ncen
%
%\author{T.Kato}\suny
%\author{C.McGrew}\suny
%\author{C.Saji}\ncen
%\author{C.W.Walter}\duke
%------------------------------------------------------------------------
%Kamioka observatory
%\author{Y.Ashie}\icrr
%\author{S.Fukuda}\icrr
%\author{Y.Fukuda}\icrr
\author{K.Abe}\icrr
%\author{J.Hosaka}\icrr
\author{Y.Hayato}\icrr
\author{T.Iida}\icrr
\author{K.Ishihara}\icrr
\author{J.Kameda}\icrr
\author{Y.Koshio}\icrr
\author{A.Minamino}\icrr
\author{C.Mitsuda}\icrr
\author{M.Miura}\icrr
% ...
\collaboration{The Super-Kamiokande Collaboration}\noaffiliation

%%%%%%%%%%%%%
\begin{abstract}
A search for Q-balls has been carried out in Super-Kamiokande II with 541.7\,days of live time. 
A neutral Q-ball passing through the detector can interact with nuclei to 
produce pions, generating a signal of successive contained pion events along a track.
No candidate for successive contained event groups has been found in Super-Kamiokande II, so we obtain upper limits on the flux of such Q-balls.
\end{abstract}
\maketitle
In the framework of some supersymmetric models, stable non-topological solitons called Q-balls~\cite{Coleman:1985ki} can be produced in the early universe and contribute to the dark matter~\cite{Affleck:1984fy,Kusenko:1997vp,Kasuya:1999wu,Kasuya:2001hg,Enqvist:2003gh}. If Q-balls are formed in the early universe and have survived until the present time, they would be part of the dark matter located in the halo of Galaxy. Q-balls are solitons with theoretically appealing roles in some dark matter scenarios and in explanations of the baryon asymmetry~\cite{Enqvist:1997si}. 

In this model~\cite{Kusenko:1997vp}, the Q-ball mass is given by
\begin{equation}
M_{Q} = \frac{4\pi \sqrt{2}}{3} {M_S} Q^{\frac{3}{4}},
  \label{eq:qball_m}
\end{equation}
and its radius is
\begin{equation}
  R_Q = \frac{1}{\sqrt{2}} {M_S}^{-1} Q^{\frac{1}{4}},
  \label{eq:qball_ra}
\end{equation}
where $M_S$ is the energy scale of SUSY breaking and Q is the Q-ball's baryon number. 
Q-balls are classified into two groups according to the properties of their interactions with matter: supersymmetric electrically charged solitons (SECS) and supersymmetric electrically neutral solitons (SENS).
Scintillator detectors may be more suitable for the detection of charged Q-balls with large energy loss in matter. Thus, we will confine our attention here to neutral Q-balls.

When a neutral Q-ball collides with a nucleon it absorbs its baryonic charge and induces the dissociation of the nucleon into free quarks. In this process, about 1\,GeV energy is released by the emission of typically two or three pions~\cite{Bakari:2000dq,Kusenko:2004yw}. This is called the KKST process~\cite{Kusenko:1997vp}.

The cross section of the interaction between neutral Q-balls and the matter is roughly estimated to be the geometrical size of the Q-ball,
\begin{equation}
  \sigma = \pi {R_Q}^2.
  \label{eq:qball_sigma}
\end{equation}
We consider cross sections between 0.02 and 200\,mb in this work.
Neutral Q-balls lose energy due to collisions with nucleons, and the rate of energy loss is~\cite{Bakari:2000dq}, 
\begin{equation}
  \frac{dE}{dx} \sim \frac{\zeta}{\lambda},
  \label{eq:eloss}
\end{equation}
where $\zeta$ =1\,GeV is the released energy in one collision and $\lambda$ is the mean interaction length. For 100\,mb cross section, the energy loss of Q-ball is estimated to be about 60\,MeV/cm. The mass of a Q-ball with this cross section is about $10^{19}$\,MeV/$c^2$. Therefore the effect of the energy loss in the Earth is expected to be negligible compared to the kinetic energy of a Q-ball.

Super-Kamiokande
is a cylindrical 50\,kton water Cherenkov detector located
 at a depth of 2,700\,m water equivalent. The water tank is optically
 separated into two regions; the inner detector~(ID) instrumented with inward facing 20\,inch diameter photomutiplier tubes~(PMT) and the outer detector~(OD) instrumented with outward facing 8\,inch PMTs.
Super-Kamiokande II (SK-II) is the second phase of Super-Kamiokande experiment and started taking data in December 2002 with 5,182\,PMTs in the ID and 1,885\,PMTs in the OD. 
%nim paper 
See~\cite{Fukuda:2002uc} for more details on the detector. The inner PMTs are instrumented with acrylic and FRP (fiber reinforced plastic) covers to avoid a chain reaction implosion. The event gate width is 1.3$\mu$sec.
 The OD is used to veto entering cosmic ray muons and to tag
 exiting charged particles. 
In this paper we report the results from a search for neutral Q-balls using 541.7\,live days of data for the SK-II period. 
%% wimp search with SK-I
Previous searches for WIMP dark matter with SK-I data have been reported in~\cite{Desai:2004pq}.
A Q-ball interacting in the detector will produce several pions which emit Cherenkov light. Thus, we can observe these pion events along the Q-ball track in the detector. 

%%%%%%% MC

We developed a Monte Carlo simulation to estimate the detection efficiency of Q-ball events. 
The direction of a Q-ball is randomly generated, and the distance between the center of the SK detector and a Q-ball track is assumed to be uniformly distributed within 50\,m.
The Q-ball velocity is generated assuming a Maxwellian distribution with 
a mean value of 270\,km/sec---a typical galactic velocity. Therefore, the velocity of Q-balls observed on the earth is determined by superposing the velocity of solar system 220\,km/sec on the Q-ball velocity in the halo~\cite{Smith:1988kw,Lewin:1995rx}. 
The pions produced in the interactions with a Q-ball and a nucleon can be determined using KNO scaling~\cite{Koba:1972mr}. The total energy of the released pions is equal to the nucleon mass. The charges of the pions are randomly generated, so that the probability of $\pi^+$, $\pi^-$, $\pi^0$ generation is equal within the constraint of the charge conservation. The kinematics of generated pions are determined by assuming the Q-ball and nucleus scatter into the Q-ball and the several pions. The number of generated Q-ball tracks is 20,000 for each cross section.

%%%% reduction
In this work we look for the signal of a neutral Q-ball interacting successively with two or more nucleons.
The main strategy of the data reduction is to reject muon events in order to observe a Q-ball signature consisting of at least two successive events in which 1\,GeV of energy is released in the form of pions in the ID.
% A Q-ball with $\sigma$=10\,mb could interact about 18\,times within 100\,$\mu$sec, since the interaction length of a Q-ball with $\beta$=10$^{-3}$ is 30\,m.
 A Q-ball with $\beta$=10$^{-3}$ covers 30\,m in 100\,$\mu$sec; assuming $\sigma$=10\,mb it interacts on average 18\,times. The details of the event reconstruction in Super-Kamiokande can be found in~\cite{Shiozawa:1999}.

The first selection criteria are: 

1-1) The number of total photoelectrons (p.e.'s) in the ID should be larger than 300 and the maximum number of p.e.'s in the ID within sliding 300\,nsec time window should be less than 10,000, which corresponds to about 60\,MeV and 2\,GeV energy for electrons, respectively. 
These cuts reject electron events from decays of cosmic ray muons, high energy through-going muon events and electrical noise events.

\begin{figure}[t]
  \includegraphics[width=0.7\linewidth]{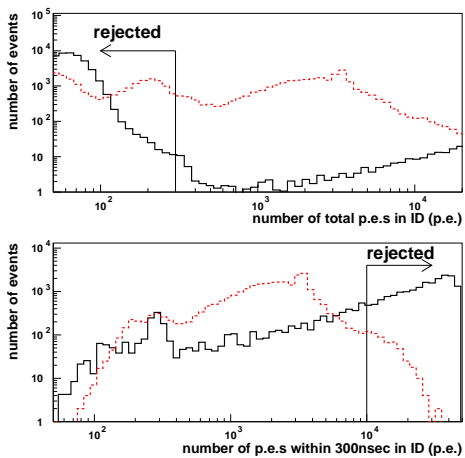} 
  \caption{The charge distributions for the data and Q-ball Monte Carlo events. The solid black line indicates the data and the dotted red line indicates the Monte Carlo events with $\sigma$=10\,mb.}
 \label{fig:1streduc}
\end{figure}

Figure~\ref{fig:1streduc} shows the charge distributions for the data and Q-ball Monte Carlo events with $\sigma$=10\,mb. The total number of p.e.'s in the ID is shown in the upper figure. The rising tail below a few hundred p.e.'s in the data has a lot of low energy events from radioisotopes and electrical noise events. On the other hand, for the Q-ball Monte Carlo events, the rising tail below 100\,p.e.'s comes from the events which do not enter the ID. The peak around 200\,p.e.'s consists of electrons from decay of charged pions. The events of neutral pions form the peaks around 4000\,p.e.'s. 
In the bottom figure we show the number of ID p.e.'s within a 300\,nsec time window for events with more than 300\,p.e.'s in the ID.. 

1-2) At least one delayed event should exist within a 100\,$\mu$sec window after the event selected by 1-1). This cut rejects single events.

All events selected by 1-1) and 1-2) are considered to be an event group. The event selected by criterion 1-1) is regarded as the first event in an event group.

After the first selection, there are still background event groups which are caused by cosmic ray muons, electrons from a decay of cosmic ray muons and electrical noise caused by large muon pulses. Background event groups still survive the above selection criteria with an approximate rate of 0.06\,/sec.

\begin{figure}[t]
  \includegraphics[width=0.7\linewidth]{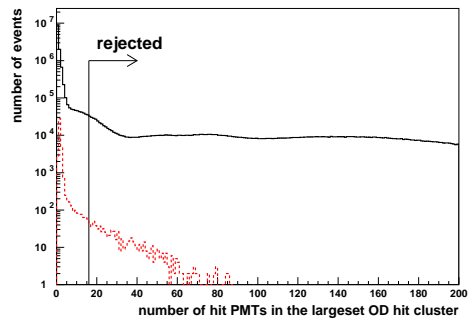} 
  \caption{The distribution of the number of hits in the largest OD cluster for the data and Q-ball Monte Carlo events. The lines are same as in Figure~\ref{fig:1streduc}.}
 \label{fig:nhitac_10mb_data}
\end{figure}

In the second reduction, we require at least two events which are entirely contained in the ID to reject cosmic ray muon events. At least two events in an event group must satisfy both of 2-1) and 2-2) criteria:

2-1) The number of hit PMTs in the largest OD hit cluster should be less than 16, in order to remove primarily cosmic ray muon events as shown in Figure~\ref{fig:nhitac_10mb_data}.

\begin{figure}[t]
  \includegraphics[width=0.7\linewidth]{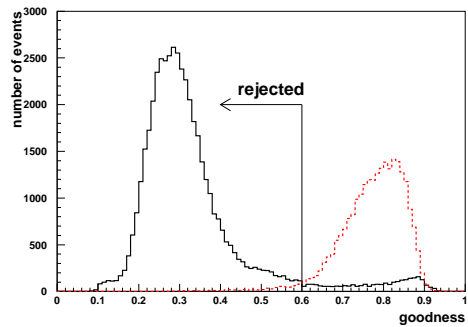} 
  \caption{The distribution of the goodness of the vertex fit for the data and the Q-ball Monte Carlo events. The lines are same as in Figure~\ref{fig:1streduc}.}
 \label{fig:goodness_10mb_data}
\end{figure}

2-2) The number of p.e.'s in the ID within 300\,nsec time window should be larger than 500 and the goodness of vertex fit should be larger than 0.6 (see Figure~\ref{fig:goodness_10mb_data}) to better reject events associated with electrical noise.

2-3) In a event group two or more events should satisfy both 2-1) and 2-2) criteria. 

%%%% bye-bye event cut

Two event groups have been found, both containing two successive events. However, they are accidental coincidence events consisting of a low energy event and a cosmic ray muon within the 1.3\,$\mu$sec trigger gate. The cosmic ray muon is separately recorded over two events since the cosmic ray muon hits at the end of the trigger gate.  These event groups are not rejected in the former reduction stages because of the absence of hit OD PMTs and the large number of total p.e.'s in the ID due to the muons.

We design the following selection criteria to reject the accidental coincidence events without losing any Q-ball Monte Carlo events.

\begin{figure}[t]
  \includegraphics[width=0.7\linewidth]{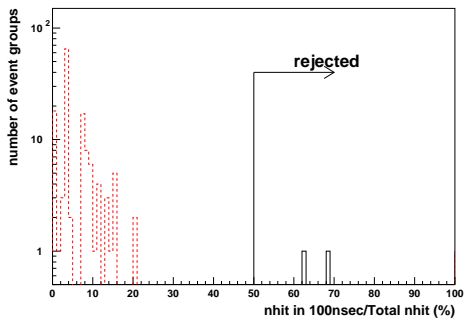} 
  \caption{The distribution of the ratio of the number of hit ID PMTs in a 100\,nsec time window at the end of the trigger gate to the total hit ID PMTs for the data and the Monte Carlo events.  The lines are same as in Figure~\ref{fig:1streduc}.}
 \label{fig:byebye}
\end{figure}

3) If an event group contains two events, for the first event, the ratio of the number of hit ID PMTs in a 100\,nsec time window at the end of the trigger gate to the total number of hit ID PMTs is required to be less than 50\,$\%$.
 The distributions of this ratio for the data and the Monte Carlo events are shown in Figure~\ref{fig:byebye}.
The events for which the ratio is more than 50\,$\%$ are defined 
to be the accidental events.

\begin{table}[t]
\begin{center}
\begin{tabular}{lcc}
\hline
\hline
Reduction step & Data & Monte Carlo (10\,mb) \\
\hline
%Trigger & - & 20000(100\,$\%$) \\
1-1)   & 16777216 & 3178(15.9\,$\%$) \\
1-2)  & 2645986 & 2867(14.3\,$\%$)\\
2-1) & 2610998 & 2865(14.3\,$\%$) \\
2-2) & 2612 & 2842(14.2\,$\%$) \\
2-3) & 2 & 2726(13.6\,$\%$) \\
Accidental event cut & 0 & 2726(13.6\,$\%$) \\
\hline
\end{tabular}
\caption{Number of event groups after each reduction step for data and the Q-ball Monte Carlo with $\sigma$=10\,mb.}
\label{table:reduc}
\end{center}
\end{table}

No group of events has survived after applying all of the selection criteria. The number of background event groups due to the accidental coincidence of 2 or more atmospheric neutrino events within 100\,$\mu$sec during the 541.7\,day exposure is estimated to be 8.6$\times$10$^{-5}$.
Table~\ref{table:reduc} shows the number of Q-ball candidates after each reduction step for data and Monte Carlo with $\sigma$=10\,mb.

Since there is no candidate event group, the upper limits on the Q-ball flux for various mean interaction lengths can be calculated from the following formula,
\begin{equation}
 Flux < \frac{N_0}{S_{eff} \Omega T}\ (90\,\% C.L.),
  \label{eq:flux}
\end{equation}
where $N_{0}=2.3$ for no observed event group, T is the detector live time (541.7\,days), and $S_{eff} \Omega$, the effective aperture, which is defined as:
\begin{equation}
S_{eff} \Omega = \epsilon \pi r^2 \times 4 \pi \ \textrm{ [m$^2$ sr]},
\end{equation}
where $\epsilon$ is the detection efficiency and {\it r} is the impact parameter, which is defined to be 50\,m in this analysis. $\epsilon$ can be estimated by Monte Carlo events separately for four different values of Q-ball cross section and is 4.5\,$\%$, 13.6\,$\%$, 14.5\,$\%$ and 14.4\,$\%$ respectively for 1\,mb, 10\,mb, 100\,mb and 200\,mb cross section.

In another theory, Q-balls could convert matter into antimatter on their surfaces~\cite{Kusenko:2004yw}. In this case, a baryon, reflected off a Q-ball as an anti-baryon, would quickly annihilate with a baryon and release about 2\,GeV of mass energy. The present selection criteria is as efficient for the Q-balls from this theory. We also find that varying the multiplicity of the generated pions from two to four does not change the detection efficiency.
 The velocity of the neutral Q-ball has uncertainty of as much as $\pm$10\,$\%$~\cite{Smith:1988kw,Lewin:1995rx}. The efficiency is not affected by this uncertainty.
Table~\ref{table:flux} summarizes the obtained flux limits. 
\begin{table}[t]
\begin{center}
\begin{tabular}{cccc}
\hline
\hline
Cross section & Detection efficiency & Flux upper limits \\
~~~~~~(mb) & ($\%$) & (cm$^{-2}$sec$^{-1}$sr$^{-1}$)\\
\hline
0.02 & 0.01 & 5.0$\times 10^{-13}$ \\
0.05 & 0.03 & 2.0$\times 10^{-13}$ \\
0.1 & 0.07 & 7.7$\times 10^{-14}$ \\
0.2 & 0.6 & 9.1$\times 10^{-15}$ \\
0.5 & 2.1 & 2.4$\times 10^{-15}$ \\
1   & 4.5 & 1.1$\times 10^{-15}$ \\
10  & 13.6 & 3.7$\times 10^{-16}$ \\
100 & 14.5 & 3.4$\times 10^{-16}$ \\
200 & 14.4 & 3.5$\times 10^{-16}$ \\
\hline
\end{tabular}
\caption{Summary of flux upper limits on neutral Q-balls for several Q-ball-nucleon interaction cross-sections.}
\label{table:flux}
\end{center}
\end{table}

\begin{figure}[t]
  \includegraphics[width=3.4in]{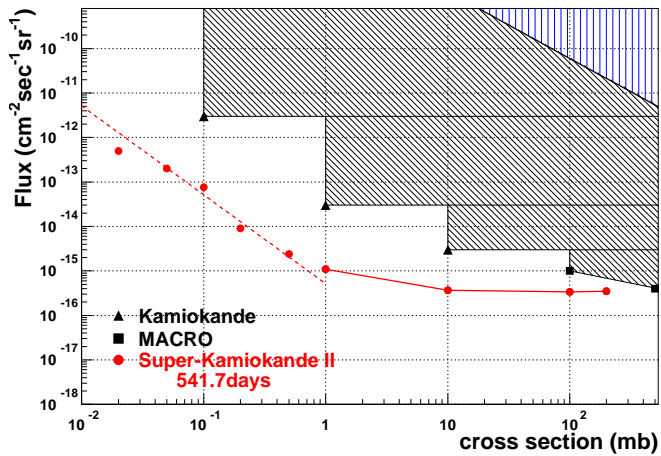} 
  \caption{90\,\% C.L. upper flux limits on the neutral Q-ball flux as a function of the Q-ball-nucleon interaction cross section. The circles show the results of this paper and the dashed line in the region below 1\,mb cross section is obtained assuming that the flux limits should be in inverse proportion to the square of cross section. }
 \label{fig:flux}
\end{figure}

Figure~\ref{fig:flux} shows the 90\,$\%$ C.L. upper limits on the neutral Q-ball flux as a function of the Q-ball cross section. The upper right hatched region is excluded because the Q-ball mass density exceeds the Galactic mass density~\cite{Arafune:2000yv}. 
The signal of the neutral Q-balls may be close to that of nucleon decays catalyzed by monopoles, so it is possible to obtain a rough estimate of the flux upper limit using the results of monopole search experiments~\cite{Arafune:2000yv}. 
The hatched region is previously excluded by the experiments which searched for monopole-catalyzed nucleon decays. The triangle points are the results from Kamiokande~\cite{Kajita:1986nb} and the square point is obtained by MACRO~\cite{Ambrosio:2002qu}. The present results, shown in circles, are the most stringent limits on neutral Q-ball flux for Q-ball cross section below 200\,mb.
We require at least 2 successive events, so the detection efficiency would be proportional to the square of the Q-ball cross section in the small cross section region. Assuming that the flux limits should be in inverse proportion to the square of cross section, the dashed line is obtained in the region below 1\,mb cross section.

In summary, we report on a search for groups of successive contained events at Super-Kamiokande II. We have found no evidence for the successive contained event groups in 541.7\,days of SK-II data. New upper limits at 90\,$\%$ C.L. for neutral Q-ball flux are obtained. These limits are the most stringent bounds on Q-ball flux for Q-ball cross section below 200\,mb. 

We gratefully acknowledge the cooperation of the Kamioka Mining and Smelting Company. The Super-Kamikande experiment has been built and operated from funding by the Japanese Ministry of Education, Culture, Sports, Science and Technology, the United States Department of Energy, and the U.S. National Science Foundation, and the National Natural Science Foundation of China.

\end{document}